\documentclass[aps,prd,draft,groupedaddress,preprintnumbers,%
              showpacs,showkeys]{revtex4}       
\usepackage[latin1]{inputenc}                    
\usepackage{graphicx}                            
\usepackage{epsfig}
\usepackage{epsf}
\usepackage{latexsym}                            
\usepackage{amsfonts}                            
\usepackage{amssymb}                             
\usepackage{amsmath}                             
\usepackage[mathscr]{eucal}                      
\usepackage{dcolumn}                             
\usepackage{bm}                                  
\usepackage{hyperref}                            

\graphicspath{{.}{Graphics/}}                    
\newcommand{\imag}{\mbox{i}}                     
\newcommand{\mps}{m_{\pi}}                       
\newcommand{\mV}{m_{\rho}}                       

\newcommand{\GmO}{{\cal G}_{M1}}                 
\newcommand{\GcT}{{\cal G}_{C2}}
\newcommand{\GeT}{{\cal G}_{E2}}

\newcommand{\be}{\begin{equation}}
\newcommand{\ee}{\end{equation}}
\newcommand{\beq}{\begin{eqnarray}}
\newcommand{\eeq}{\end{eqnarray}}
\begin{document}

\preprint{} \title[Nucleon-Delta transition]{The nucleon to
  Delta electromagnetic transition form factors in lattice QCD}
\author{C.~Alexandrou}\affiliation{Department of Physics, University
  of Cyprus, CY-1678, Cyprus}
\email{alexand@ucy.ac.cy}
\author{G.~Koutsou}\affiliation{Department of Physics, University
  of Cyprus, CY-1678, Cyprus}
\author{H.~Neff}\affiliation{Center for Computational Science,
 University College London, 20 Gordon Str., London WC1, UK}
\author{J.W.~Negele}\affiliation{Center for Theoretical Physics,
  Massachusetts Institute of Technology, Cambridge, MA 02139, USA}
\author{W.~Schroers}
\affiliation{Department of Physics, Center for Theoretical Sciences,
National Taiwan University, Taipei 10617, Taiwan}
  \email{Wolfram.Schroers@Field-theory.org}
\author{A.~Tsapalis}\affiliation{University of Athens, Institute of
  Accelerating Systems and Applications, Athens, Greece}
\date{\today}

\begin{abstract}
The electromagnetic nucleon to $\Delta$ transition form factors
 are evaluated using
two degenerate flavors of dynamical Wilson fermions 
and   using dynamical sea
staggered fermions with domain wall valence quarks.
  The two subdominant quadrupole form factors 
are evaluated for the first time in full QCD to sufficient accuracy 
to exclude a zero value,  which is  taken as a signal  for deformation 
in the nucleon-$\Delta$ 
system. For the Coulomb quadrupole form factor the  unquenched 
results show   
 deviations from the quenched results at low $q^2$   
 bringing dynamical lattice results 
closer to experiment, thereby 
confirming the importance of pion cloud contributions on this quantity.

 \end{abstract}


\maketitle

%
%

\section{\label{sec:Introduction}Introduction}
Despite several decades of scrutiny, the intrinsic shape of the nucleon, 
a fundamental building block of our world, is still not
fully   resolved~\cite{Isgur,Capstick:1990}. Although the nucleon is
experimentally easily accessible in exclusive and inclusive scattering 
its spectroscopic quadrupole moment is zero since it
has spin $J=1/2$. However, this does not mean that the nucleon is spherically 
symmetric, since it can have an intrinsic deformation. Deformation is a
 common phenomenon in nuclear and atomic physics.
 Quantum mechanically, a
multiphoton coincidence experiment could  determine that a $J=0$ 
ground
state of a diatomic molecule or
nucleus has a deformed shape.  However, usually in
electromagnetic probes of microscopic systems, we are constrained to make
measurements associated with one-photon exchange, corresponding to a
matrix element of a one-body operator. In the case of a diatomic molecule,
the  one-body  charge density of the $J=0$ state is spherically symmetric,
and cannot reveal the deformation that is present in the system.
For $J>1/2$, however, when a nuclear or atomic system is well
approximated by a deformed intrinsic state, it is still possible to
observe its deformation using a one-body electromagnetic operator.  In this
case,  the quadrupole moment is non-zero
 in the 
laboratory frame if the state is deformed. For collective
rotation of the deformed intrinsic state~ ~\cite{Bohr},
the relation between the
spectroscopic quadrupole moment, $Q$, measured in the laboratory frame and
the intrinsic quadrupole moment, $Q_0$, in the body-fixed intrinsic frame
is given by
\be
Q=\frac{3K^2-J(J+1)}{(J+1)(2J+3)} Q_0\quad,
\label{quadrupole}
\ee
where $J$ is the total angular momentum of the system in the laboratory frame,
 $K$ is the
projection of $J$ onto the z-axis of the body-fixed intrinsic frame, and we
have considered the sub-state with azimuthal quantum number $M=J$.  In the
previous example of the $J=0$ diatomic molecule, although $Q_0 \ne 0$, 
Eq.~(\ref{quadrupole})
yields $Q=0$ so that the deformation is unobservable.
Similarly, in the case of a nucleon with $J=1/2$, $Q$ is zero although $Q_0$
may not 
be~\cite{Alexandrou:2002prd, Alexandrou:2003Cairns}.
Since the $\Delta$ has $J=3/2$ a non-zero quadrupole moment $Q$ 
can be measured~\cite{Alexandrou:DD2007}. 
The electric and Coulomb transition amplitudes E2 and C2
 between the $J=1/2$ nucleon and
its $J=3/2$ resonance have the same property of revealing the presence of
deformation in the nucleon, the $\Delta$, or both.
Therefore,   as  in experiment, in this work we detect deformation
by measuring E2 and C2.

 In  recent years, we have seen tremendous progress in experimental
measurements of the subdominant quadrupole 
amplitudes~\cite{Blanpied:1996,Beck:2000,Mertz:2001,MAMI2:2001,
Joo:2001,Sparveris:2005,Stave:2006,Sparveris:2007}, yielding very accurate
results  particularly at low $q^2$. For 
a recent review of the experimental situation 
see Ref.~\cite{Papanicolas:2006athens}.
These accurate measurements have motivated several 
recent theoretical studies
both in lattice QCD~\cite{Alexandrou:EM_N2D2004prl, Alexandrou:EM_N2D2004, Alexandrou:EM_N2D2005}  
and in chiral effective theories~\cite{Marc:2006athens,Hemmert:2006athens}.
On the lattice, hadron deformation can also be studied by
investigating directly the charge distribution
 using density-density correlators~\cite{Alexandrou:2002prd,Alexandrou:2003prd, Alexandrou:den_den2003}, and
 it was shown
 that  the rho meson is deformed~\cite{Alexandrou:den_den2007}. The $\Delta$ 
shows deviations from a
spherical shape albeit with large statistical errors. The issue of deformation
of the $\Delta$ using density-density correlators is under study with
improved lattice techniques~\cite{Alexandrou:den_den2005, 
Alexandrou:den_den2006}.    
 For concise reviews
see~\cite{Alexandrou:2006athens,Alexandrou:2003Cairns}.

The focus of this work is a calculation of  the
nucleon - $\Delta$ electromagnetic transition form factors within QCD.
We use two different types of simulations:
The first   uses two degenerate flavors of dynamical Wilson fermions
and  the second a hybrid action.
 The 
hybrid action uses dynamical staggered sea quarks, with
  two degenerate light quarks  and  one fixed to the mass of the strange
quark. These dynamical quark configurations are
 produced by the MILC collaboration~\cite{Bernard:2001MILC}
 and represent a state-of-the-art
simulation of the QCD vacuum  with  the three lightest
flavors of quarks taken into account.  For the light valence
quarks we use two degenerate domain wall fermions.
This approach has been  used  successfully  to evaluate
nucleon structure functions~\cite{Negele:2004iu,Renner:2004,
  Hagler:2004er,Edwards:2005ym,Edwards:2005kw,Edwards:2006qx,Hagler:2007}
 and the N to $\Delta$ 
axial-vector
form factors~\cite{Alexandrou:axial_N2D2007prl, Alexandrou:axial2007prd,
Alexandrou:axial_N2D2006,Alexandrou:axial2007}.
In this work, we compare results  calculated using the hybrid action to
 the results obtained 
using two degenerate flavors of dynamical Wilson fermions. Given the
different lattice artifacts involved in the two approaches, agreement
between them provides a consistency check of our lattice methodology.
Comparison between the dynamical  results and our quenched 
results~\cite{Alexandrou:EM_N2D2004prl,Alexandrou:JLab2007} probes 
pion cloud contributions.

The first lattice study of the N to $\Delta$ electromagnetic transition form 
factors~\cite{Leinweber:1993} was carried out in the quenched approximation
 with  limited statistics. Although the mean value
was negative, the statistical errors 
on the suppressed quadrupole amplitudes were large and  
a zero value could not be excluded. 
 Nonetheless, this pioneering work set up the
 methodology for a more elaborate study  that would soon become feasible
once sufficient computing resources  became available. Using
the  approach  of Ref.~\cite{Leinweber:1993} we evaluated the
transition form factors using
quenched and dynamical Wilson quarks going to smaller quark masses
  than those of Ref.~\cite{Leinweber:1993} but only at the
lowest $q^2$-value allowed on our 
lattices~\cite{Alexandrou:EM_N2D2004prd}.
However, although we increased  statistics, 
 the  quadrupole form factors
still had large errors. Constructing optimized sources
 that led to a large  sample of statistically independent 
measurements for a given $q^2$-value and carrying out
sequential inversions through the source instead of through the current,
 we were  finally able to obtain both quadrupole  form factors
 to sufficient accuracy
for a range of $q^2$-values. This calculation, carried out in the
quenched approximation, confirmed
a non-zero value  with  the correct sign for both the quadrupole
amplitudes~\cite{Alexandrou:EM_N2D2004prl}. Using this new methodology we
extend in this work the calculation to the unquenched case.   Initial
 unquenched 
results have been reported in 
Refs.~\cite{Alexandrou:EM_N2D2003, Alexandrou:EM_N2D2004, Alexandrou:EM_N2D2005}.

Lattice calculations at the physical quark mass are  currently  prohibitively
expensive. Recently, progress in both hardware performance and
algorithms~\cite{Schroers:2007qf} has extended the range of
accessible quark masses to lower masses,
 bringing lattice calculations into the region where they 
can be extrapolated using
chiral perturbation theory.
In  this work, we use pions with masses as low as about 350 MeV.
 In a previous work~\cite{Alexandrou:EM_N2D2004prl}, we found
a discrepancy for  the CMR ratio between 
the quenched lattice results and experiment at low values of $q^2$.
 On the other hand,  for larger $q^2$-values,
 the quenched results and experiment were in quantitative agreement. 
Our current results for CMR 
show deviations between unquenched and quenched results 
at the lowest momentum transfer. In fact, the previously observed discrepancy 
is reduced by our new unquenched results. This could indicate 
that pion cloud effects are quite important at low $q^2$ and at 
light pion masses, as also discussed in the framework of chiral perturbation 
theory~\cite{Pascalutsa:2005}.   
With pion masses in the range of 300 MeV as planned in the future 
one hopes to make progress  in reliable chiral extrapolations to
the physical regime.
Such progress has been demonstrated recently
in the extrapolation of   the nucleon axial
coupling
within chiral effective theory~\cite{Edwards:2005ym,Khan:2006de}.

This paper is organized as follows: Section~\ref{sec:NtoDelta matrix element}
gives the decomposition
of the N to $\Delta$ 
matrix element on the hadronic level, Section~\ref{sec:latt-matrix}
and Section~\ref{sec:plateaus}
detail the lattice analysis  and outline
our strategy for extracting observables.
Section~\ref{sec:results-nucleon-tran} contains our results for the
transition form factors.
 Finally, Section~\ref{sec:conclusions} contains our
conclusions and an outlook for further studies we intend to perform in
the field.

\section{\label{sec:NtoDelta matrix element}N to Delta matrix element}

The electromagnetic transition matrix element can be expressed in terms
of the  
three Sachs form factors~\cite{Jones:1973},
\begin{equation}
  \label{eq:trans-mat}
  \langle\Delta({\bf p}',s')\vert j_\mu\vert N({\bf p},s)\rangle =
  \imag\,\left(\frac{2}{3}\right)^{1/2} \biggl(\frac{m_{\Delta}\; m_N}{E_{\Delta}({\bf p}^\prime)\;E_N({\bf p})}\biggr)^{1/2}\bar{u}_\sigma(\bf{p}',s')
  {\cal O}_{\sigma\mu} u(\bf{p},s) \,,
\end{equation}
with the Lorentz-structure
\begin{equation}
  \label{eq:lorentz-op}
  {\cal O}_{\sigma\mu} = \GmO(q^2) K_{\sigma\mu}^{M1} + \GeT(q^2)
  K_{\sigma\mu}^{E2} + \GcT(q^2) K_{\sigma\mu}^{C2}\, .
\end{equation}
The kinematic prefactors in Euclidean space are given by
\begin{eqnarray}
  \label{eq:lorentz-op-prefac}
  K_{\sigma\mu}^{M1} &=& - \frac{3}{(m_\Delta+m_N)^2+Q^2}
  \frac{m_\Delta+m_N}{2m_N}\,\imag\,
  \varepsilon_{\sigma\mu\alpha\beta}\, p^\alpha 
  {p'}^\beta\,, \nonumber \\
  K_{\sigma\mu}^{E2} &=&  -K_{\sigma\mu}^{M1} + 6\, \Omega^{-1}(Q^2)
  \frac{m_\Delta+m_N}{2m_N} 2\,\imag\, \gamma_5\,
  \varepsilon_{\sigma\lambda\alpha\beta}\, p^\alpha {p'}^\beta
  \varepsilon_{\mu}^{\phantom{\mu}\lambda\gamma\delta}\, p_\gamma
  {p'}_\delta\,, \nonumber\\
  K_{\sigma\mu}^{C2} &=& -6\, \Omega^{-1}(q^2)
  \frac{m_\Delta+m_N}{2m_N}\, \imag\, \gamma_5\, q_\sigma \left( q^2
    (p+p')_\mu - q \cdot (p+p') q_\mu \right)\,,
\end{eqnarray}
where $p(s)$ and $p'(s')$ denote initial and final momenta (spins),  
$q^2\equiv(p'-p)^2$,  and
$ u_\sigma (p',s')$ is a spin-vector in the Rarita-Schwinger formalism.
We define $\Omega(Q^2) =  \left[(m_\Delta+m_N)^2+Q^2\right]
\left [(m_\Delta-m_N)^2+Q^2\right]$ 
and ${\bf Q}={\bf q}$, $Q^4=iq^0$ is the lattice
 momentum transfer giving $Q^2=-q^2$. 
The Rarita-Schwinger spin sum for the $\Delta$  in Euclidean space is given by
\be
\sum_s u_\sigma(p,s)\bar{u}_\tau(p,s) = \frac{-i\gamma . p+m_\Delta}{2m_\Delta} \left[\delta_{\sigma\;\tau} +\frac{2p_\sigma p_\tau}{3m_\Delta^2}
-i\frac{p_\sigma\gamma_{\tau}-p_\tau\gamma_{\sigma}}{3m_\Delta}
-\frac{1}{3}\gamma_\sigma \gamma_\tau \right] ,
\ee 
and the Dirac spin sum 
\be
\sum_s u(p,s)\bar{u}(p,s) =\frac{-i\gamma .p+m_N}{2m_N} \quad.
\ee

The magnetic dipole form factor, $\GmO$, is the
dominant form factor.  The electric and Coulomb
quadrupole form factors, $\GeT$, and
$\GcT$ are subdominant.
 They are scalar
functions depending  on the  momentum transfer  squared.
 On the lattice we can only access space-like 
$q^2$ and therefore  $Q^2>0$.
The reason for
using the Sachs parametrization in a lattice computation,
as pointed out in Ref.~\cite{Leinweber:1993}, is that
the Sachs form factors  do not depend strongly on the difference
between the
nucleon and the $\Delta$ mass.

It is customary to quote the ratios of the electric and Coulomb quadrupole
amplitudes to the magnetic dipole amplitude, $R_{EM}$ (EMR) and $R_{SM}$ (CMR).
In
 the  rest frame of the $\Delta$, 
they are given by~\cite{Jones:1973,Gellas:1999}
\begin{eqnarray}
  \label{eq:ratios-def}
  R_{EM} &=& \displaystyle\nonumber -\frac{\GeT(q^2)}{\GmO(q^2)}\,, \\
  R_{SM} &=& \displaystyle -\frac{\vert\vec{q}\vert}{2m_\Delta}
  \frac{\GcT(q^2)}{\GmO(q^2)}\,.
\end{eqnarray}
Deformation of the nucleon-$\Delta$ system
  will produce non-zero values of 
 $R_{EM}$ and $R_{SM}$.   Using accurate measurements of the longitudinal-transverse 
response function
at $q^2=-0.127$~GeV$^2$ it has 
been shown that a spherical   nucleon - $\Delta$ system is inconsistent
with the experimental results~\cite{Papanicolas:2003,Sparveris:2005}.

%
%

\section{\label{sec:latt-matrix}Lattice matrix element}

The   most computationally  demanding part in this evaluation is the
calculation
 of the 
 three-point correlation function  given by 
\be
\langle G^{\Delta j^\mu N}_{\sigma} 
(t_2, t_1 ; {\bf p}^{\;\prime}, {\bf p}; \Gamma) \rangle =
\sum_{{\bf x}_2, \;{\bf x}_1}
\exp(-i {\bf p}^{\;\prime} \cdot {\bf x}_2 )  
\exp(+i ({\bf p}^{\;\prime} -{\bf p}) \cdot {\bf x}_1 ) \;  
\Gamma^{\beta \alpha}
\langle \;\Omega \; | \; T\left[\chi^{\alpha}_{\sigma}({\bf x}_2,t_2) 
j^{\mu}({\bf x}_1,t_1) \bar{\chi}^{\beta} ({\bf 0},0) \right]
\; | \;\Omega \;\rangle   \; ,
\label{Delta N}
\ee
where an initial state with the quantum numbers of the  
nucleon is created 
at time zero and the final state with the quantum numbers of the
$\Delta$ is annihilated at a later time $t_2$. The photon couples
to one of the quarks in the nucleon at an intermediate time $t_1$
producing a $\Delta$.
For the spin-$\frac{1}{2}$ source, $\chi ^p ({\bf x},t)$, and the 
spin-$\frac{3}{2}$ source, 
$\chi ^{\Delta^{+}}_\sigma  ({\bf x},t)$, we use the interpolating fields 
\be
\chi ^p (x) = \epsilon^{a b c}\; \left[ u^{T\; a}(x)\; C \gamma_5
d^b(x) \right]\; u^c(x) \quad,
\ee
\be
\chi ^{\Delta^{+}}_\sigma  (x) = \frac{1}{\sqrt{3}} \epsilon^{a b c} \Big
\lbrace
2 \left[ u^{T a}(x)\; C \gamma_\sigma d^b(x) \right]u^c(x) \;
+\; \left[ u^{T a}(x)\; C \gamma_\sigma u^b(x) \right]d^c(x) \Big \rbrace \quad,
\ee
and for the projection matrices for the Dirac indices
\be
\Gamma_i = \frac{1}{2}
\left(\begin{array}{cc} \sigma_i & 0 \\ 0 & 0 \end{array}
\right) \;\;, \;\;\;\;
\Gamma_4 = \frac{1}{2}
\left(\begin{array}{cc} I & 0 \\ 0 & 0 \end{array}
\right) \;\; .
\ee

For large Euclidean time separations 
$t_2 -t_1 \gg 1$ and $t_1 \gg 1$,
the time dependence and field normalization constants  
cancel in the following ratio
\begin{eqnarray}
  \label{eq:ratio}
  R_\sigma(t_2,t_1,{\bf p}',{\bf p},\Gamma,\mu) &=& \displaystyle
  \frac{G_\sigma^{\Delta j_\mu N}(t_2,t_1,{\bf p}',{\bf p},\Gamma)}
  {G_{ii}^{\Delta\Delta}(t_2,{\bf p}',\Gamma_4)}
  \left[ \frac{G^{NN}(t_2-t_1,{\bf p},\Gamma_4)
      G_{ii}^{\Delta\Delta}(t_1,{\bf p}',\Gamma_4)
      G_{ii}^{\Delta\Delta}(t_2,{\bf p}',\Gamma_4)}
    {G_{ii}^{\Delta\Delta}(t_2-t_1,{\bf p}',\Gamma_4)
      G^{NN}(t_1,{\bf p},\Gamma_4)
      G^{NN}(t_2,{\bf p},\Gamma_4)}\right]^{1/2}\nonumber \\
  && \displaystyle \stackrel{\stackrel{t_2-t_1\gg 1}{t_1\gg
  1}}{\to} \Pi_\sigma({\bf p}',{\bf p},\Gamma,\mu)\, ,
\label{R-ratio}
\end{eqnarray}
where $ G^{NN}$ and $ G^{\Delta \Delta}_{ij}$ are the 
 nucleon and $\Delta$ two-point functions given  respectively by
\beq
\langle G^{NN} (t, {\bf p} ; \Gamma) \rangle &=&  \sum_{{\bf x}}
e^{-i {\bf p} \cdot {\bf x} } \; \Gamma^{\beta \alpha}\;\langle \Omega |\;T\;\chi^{\alpha}({\bf x},t) 
 \bar{\chi}^{\beta} ({\bf 0},0)  
\; |  \Omega\;\rangle \nonumber \\
\langle G^{\Delta \Delta}_{\sigma\tau} (t, {\bf p} ; \Gamma) \rangle &=&  \sum_{{\bf x}}
e^{-i {\bf p} \cdot {\bf x} } \; \Gamma^{\beta \alpha}\;\langle \Omega |\;T\;\chi^{\alpha}_{\sigma}({\bf x},t) 
 \bar{\chi}^{\beta}_{\tau} ({\bf 0},0)  
\; | \Omega \;\rangle \quad. \nonumber \\
&\>&
\label{NN}
\eeq
For the Wilson fermionic action we use the lattice conserved   electromagnetic current,   $j^\mu (x)$,
given by
\be
j_\mu (x) = \sum_{f} Q_{f} \kappa_{f} \lbrace
\bar{\psi}^{f} (x + \hat{\mu})(1 + \gamma_\mu)
U^{\mu \dagger} (x) \psi^{f} (x) 
-\bar{\psi}^{f} (x)(1 - \gamma_\mu)
U^{\mu} (x) \psi^{f} (x + \hat{\mu}) \rbrace
\ee
symmetrized on site $x$ by taking
$
j^\mu (x) \rightarrow \left[ j^\mu (x) + j^\mu (x - \hat \mu) \right]/ 2
,$ where $Q_f$ is the charge of a quark of flavor $f$ and $\kappa_f$
is its hopping parameter.
For domain wall fermions, we use the local four-dimensional 
electromagnetic current
$\sum_{f} Q_{f}  \lbrace
\bar{\psi}^{f} (x)\gamma_\mu\psi^f(x) \rbrace$. This is not conserved and
therefore, to relate lattice and continuum results,
 we need the renormalization constants $Z_V$, which are known. 
Throughout this work we  { \bf choose} a frame where the $\Delta$ is at rest.
For these
kinematics we have ${\bf q}=-{\bf p}$, ${\bf p}'={\bf 0}$. At
sufficiently large time separations, $t_2-t_1$ and
$t_1$,  $R_\sigma$ becomes independent of time yielding the
desired ratio $\Pi_\sigma$.

 At the hadronic
level, with the inclusion of complete sets of baryonic states and
the use of Dirac and Rarita-Schwinger spinors the ratio of
Eq.~(\ref{eq:ratio})
 leads to the relations

\be \label{P1}
\Pi_{\sigma}({\bf 0},-{\bf q}\; ; \Gamma_4 ;\mu)= i A \epsilon^{\sigma
4\mu j} p^j {\cal G}_{M1}(Q^2) \quad , 
\ee
 \beq \label{P2}
\Pi_{\sigma}({\bf 0},-{\bf q}\; ; \Gamma_k ;j)&=&  A \Biggl\{\frac{1}{2}
\left( p_\sigma\delta_{kj}-p_k\delta_{\sigma j}\right) {\cal G}_{M1}(Q^2) 
-\biggl[ \frac{3}{2}\left(
p_\sigma\delta_{kj}+p_k\delta_{\sigma j} \right )
            - \frac{3 p_\sigma p_k p_j}{{\bf p}^2} \biggr]  {\cal G}_{E2}(Q^2)
\nonumber \\
 & \>& \hspace*{1cm}- \frac{(E_N-m_\Delta)}{2 m_\Delta}\>p_j \>
\biggl(\delta_{\sigma k}-\frac{3 p_\sigma p_k}{{\bf p}^2} \biggr)
{\cal G}_{C2}(Q^2) \Biggr\} 
\eeq
for $j=1,2,3$ and 
\be \label{P3}
\Pi_{\sigma}({\bf 0},-{\bf q}\; ; \Gamma_k ;4)= i
B \Bigl(\delta_{\sigma k}-\frac{3 p_\sigma p_k}{{\bf p}^2} \Bigr)
{\cal G}_{C2}(Q^2) \quad,
\ee
 with the kinematical coefficients 
\be 
A=\sqrt{\frac{2}{3}}\;\frac{m_\Delta+m_N}{4m_N
E_N}\;\sqrt{\frac{E_N}{E_N+m_N}} , \hspace{0.3cm} B = \frac{{\bf
p}^2}{2 m_\Delta} A.
\nonumber 
\ee
 A convenient method for the
evaluation of three-point functions is the {\it sequential
inversion through the sink}.
This requires fixing the hadronic state at $t_2$ to be the
$\Delta$ with fixed vector index $\sigma$. The
projection matrices $\Gamma $ at the sink are also fixed, but
the operator inserted at any time $t_1$ can be left arbitrary.
Therefore, with one sequential inversion one can  evaluate  the
three-point function for a large set of  lattice
momentum transfer  values  ${\bf q}$, any current
direction $\mu$  and any operator
insertions at any intermediate time $t_1$. 
One then looks for
 a plateau as a function
of $t_1$  that determines $\Pi_\sigma$.
Eqs.~(\ref{P1}-\ref{P3})
constitute a system of equations for the form factors ${\cal
G}_{M1}, {\cal G}_{E2}$ and ${\cal G}_{C2}$ at each value of
$Q^2$. It must be noted that while three independent measurements
of $\Pi_{\sigma}({\bf q}\; ; \Gamma ;\mu)$ suffice for the
determination of the form factors, increasing the combinations of
$\mu $ and photon momentum ${\bf q}$ which are
measured improves the statistical accuracy of the form factors.
With that goal in mind, we observe that Eqs.~(\ref{P1}-\ref{P3}) are
identically zero for several values of ${\bf q} = - {\bf p}$, e.g.
Eq.~(\ref{P1}) is zero when $j=\sigma$ or $j=\mu$. Furthermore, for a
given selection of $\sigma$, not all lattice rotations of ${\bf
q}$ are giving non-zero contributions. We therefore search for the
linear combinations of Eqs.~(\ref{P1}-\ref{P3}) which maximize the number of
non-zero ${\bf q}$ contributions in a lattice rotationally
invariant fashion and construct the following optimal combinations
\be \label{S1} 
S_1({\bf q};\mu)= \sum_{\sigma=1}^3\Pi_\sigma({\bf 0},-{\bf
q}\; ;
\Gamma_4 ;\mu) =  i A \biggl\{ (p_2-p_3)\delta_{1,\mu} 
 + (p_3-p_1)\delta_{2,\mu} + (p_1-p_2)\delta_{3,\mu} \biggr\}
{\cal G}_{M1}(Q^2) 
\ee
\beq \label{S2} 
S_2({\bf q};\mu)
=\sum_{\sigma\neq k=1}^{3} \Pi_\sigma({\bf 0},-{\bf q}\; ;
 \Gamma_k ;\mu) & =&
-3 A \Biggl\{  \bigl( (p_2+p_3)\delta_{1,\mu}
 + (p_3+p_1)\delta_{2,\mu} +
(p_1+p_2)\delta_{3,\mu} \bigr ) {\cal G}_{E2}(Q^2) \nonumber \\ 
 &\>& \hspace*{1cm}- 2 \frac{p_\mu}{{\bf p}^2}\bigl( p_1 p_2 + p_1 p_3 +
p_2 p_3 \bigr) 
\left[ {\cal G}_{E2}(Q^2) + \frac{E_N-m_\Delta}{2 m_\Delta} 
{\cal G}_{C2}(Q^2)
\right] \Biggr\} \; , 
\eeq
 for   $\mu = 1,2,3 $.
For $\mu =4$ we have 
\be \label{S2b} 
S_2({\bf q};\mu = 4)= 
\frac{-i\; 6\;B}{{\bf p}^2} (p_1 p_2 + p_1 p_3 + p_2 p_3)
{\cal G}_{C2}(Q^2) \> . 
\ee 
The three-point functions involved in $S_1$ and $S_2$
require one sequential inversion if one uses the appropriate linear combination
directly in the  construction of the $\Delta$ sink.
  Thus, with two inversions we obtain
the maximal number of
lattice determinations of the form factors for all the allowed
 lattice photon momenta ${\bf q}$. The $S_1$-type matrix
element determines ${\cal G}_{M1}$ while the quadrupole form factors are
 extracted from the $S_2$-type matrix element. The method is
clearly superior to the method used in 
Refs.~\cite{Leinweber:1993, Alexandrou:EM_N2D2004prd}
since the same CPU cost allows the evaluation of the form factors
at all
 $Q^2$. It should be noted
that the evaluation using the $S_2$-type sink
 does not determine  ${\cal G}_{C2}$ at the lowest allowed photon momentum
${\bf q} = (1,0,0)\,2\pi/La$ 
(and at the equivalent momentum in the other two directions).
For this reason we use, in addition, the combination
\be
S_3({\bf q};\mu)= \Pi_3({\bf 0},-{\bf q}; \Gamma_3 ;\mu) 
-\bigl(\Pi_1({\bf 0},-{\bf q}; \Gamma_1 ;\mu)
+\Pi_({\bf 0},-{\bf 0},-{\bf q}; \Gamma_2 ;\mu)\bigr)/2 
\label{S3}
\ee
to get  ${\cal G}_{C2}$ for the values of ${\bf q}$ for which
the $S_2$-type sink vanishes. 
This linear combination gives
\beq
S_3({\bf q};\mu)= 
-\frac{3\; A}{2}\; p_\mu \; \Biggl[ 
3\; \biggl( \delta_{\mu,3}-\frac{p_3^2}{{\bf p}^2} \biggr)\; {\cal G}_{E2}(Q^2) 
+\frac{E_N-m_\Delta}{2 m_\Delta} \; \biggl(1 - 3\;\frac{p_3^2}{{\bf p}^2} \biggr) \; {\cal G}_{C2}(Q^2) 
\Biggr]
\eeq
for $\mu = 1,2,3$ and
\beq
S_3({\bf q};\mu = 4)= 
\frac{3\; i\;B}{2} \biggl( 1-3\; \frac{p_3^2}{{\bf p}^2}  \biggr) \;
{\cal G}_{C2}(Q^2) \> . 
\eeq

The full set of
lattice measurements for the type $S_1$, $S_2$ and $S_3$ matrix elements
for all contributing values of $\mu$ and ${\bf q}$ at a given $Q^2$
are analyzed simultaneously. 
We denote by $P({\bf q};\mu)$ the lattice
measurements for the ratios $R_\sigma$ using the three-point
function constructed using 
the sink-types described in Eqs.~(\ref{S1}-\ref{S3}) and with $w_k$ their
statistical error. If we denote the solution vector by
$F =  \left(\begin{array}{c} {\cal G}_{M1} \\
                                   {\cal G}_{E2} \\
                                   {\cal G}_{C2} \end{array}\right)$
we are led to the overcomplete set of equations 
\be  P({\bf
q};\mu)= D({\bf q};\mu)\cdot F(Q^2) \, .
  \label{eq:overdet-set} 
\ee
The kinematical prefactors, $ D(\bf{q},\mu)$, are known
analytically and we have inserted the continuum expressions as
detailed in  Eqs.~\eqref{eq:lorentz-op} and \eqref{eq:lorentz-op-prefac}. 
As already pointed out, we have lattice measurements 
for all possible vector current components $\mu$ and photon momentum 
vectors $q$ which contribute to a given value of $Q^2$. For $N$ such 
measurements, the matrix $D({\bf q};\mu)$ is an 
$N\times 3$ matrix of kinematical 
coefficients.
 The solution
vector is determined from the minimization of the total $\chi^2$
\be \chi^2=\sum_{k=1}^{N}
\Biggl(\frac{\sum_{j=1}^3 D_{kj}F_j-P_k}{w_k}\Biggr)^2 .
\label{chi2}
\ee
Defining $D^{\prime}_{kj} = D_{kj}/w_k$ and $ P^{\prime}_k = P_k
/w_k $, $(k=1,2,..N, j=1,2,3)$ the solution  is obtained using
 the singular value decomposition of
the matrix $D^{\prime}_{kj}({\bf q};\mu)$ 
\be D^{\prime} = U \cdot
diag(\lambda_1,\lambda_2,\lambda_3) \cdot V^T \quad,
\ee 
where $U$ is an
$N \times 3$ matrix, $V$ is a $3 \times 3$ matrix and the
$\lambda_j$ are the non-negative, singular values of $D^{\prime}$.
 The form factors are therefore given by
 \be
F = V \cdot diag(1/\lambda_1,1/\lambda_2,1/\lambda_3) \cdot (U^T
\cdot P^{\prime}) . 
\ee 
This strategy has been developed
in~\cite{Hagler:2003jd} and also applied
in~\cite{Gockeler:2003jf} and subsequent publications.
The errors $w_k$ in the lattice
measurements and the final error on the form factors are
determined from the jackknife procedure.
Having decided to use sequential inversions through the sink, the
sink-source separation must be kept fixed. One would like to use the 
smallest sink-source separation that guarantees that excited state 
contributions are negligible. In our previous work we found
that a sink-source separation,$t_2$, of  about 5~GeV$^{-1}$ is sufficient.
We check in the next Section that this is also sufficient for this calculation by comparing
to the results obtained when we increase $t_2$ by about 25\%.

\begin{table}
\begin{center}
\begin{tabular}{ccccccc}
\hline \hline\multicolumn{7}{c}
{Wilson fermions}\\
\hline V& \# of confs & $\kappa$ & $m_\pi$~(GeV)&$m_\pi/m_\rho$ & $m_N$~(GeV) & $m_\Delta$ (GeV)\\ \hline
\multicolumn{7}{c}{Quenched, $\beta=6.0,~~a^{-1}=2.14(6)$~GeV} \\
\hline  
$32^3\times 64$& 200 & 0.1554 &0.563(4)& 0.645(9)  & 1.267(11) & 1.470(15)\\
$32^3\times 64$& 200 & 0.1558 &0.490(4)& 0.587(12) & 1.190(13) & 1.425(16)\\
$32^3\times 64$& 200 & 0.1562 &0.411(4)& 0.503(23) & 1.109(13) & 1.382(19)\\
 &  &$\kappa_c$ =0.1571& 0.& & 0.938(9) &\\
  \hline
  \multicolumn{7}{c}{Unquenched, $\beta=5.6,~~a^{-1} = 2.56(10)$~GeV}\\
  \hline
  $24^3\times 40$&185~\cite{TchiL} &0.1575 &0.691(8)&0.701(9) &1.485(18) & 1.687(15)\\
  $24^3\times 40$&157~\cite{TchiL} &0.1580 &0.509(8)&0.566(12) &1.280(26) & 1.559(19)\\
 $24^3\times 32$& 200~\cite{Urbach} & 0.15825 &0.384(8)& 0.453(27)&1.083(18) & 1.395(18)\\
 &  &$\kappa_c$ = 0.1585 &0. & &0.938(33) &\\
  \hline\hline
\end{tabular}
\end{center}
\caption{Parameters and number of gauge field configurations used
  in the calculation using  Wilson fermions.}
\label{Table:Wilson params}
\end{table}
\begin{table}[htbp]
  \centering
  \begin{tabular}{c*{8}{|c}}
    \hline\hline $V$ & \# of confs & $L_S m_\pi$ (fm) & $(am_{\mbox{\tiny
        u,d/s}})^{\mbox{\tiny Asqtad}}$ & $(am_{\mbox{\tiny
        u,d}})^{\mbox{\tiny DWF}}$ & $\mps$ (GeV) & $\mps / \mV$ &
    $m_N$ (GeV) & $m_\Delta$ (GeV) \\ \hline
    $20^3\times 32$ & 150 & 2.5 & 0.03/0.05 & 0.0478 & 0.606(2) &
        0.588(7) & 1.329(9) & 1.662(21) \\
  $20^3\times 32$   & 150 & 2.5 & 0.02/0.05 & 0.0313 & 0.502(4) & 0.530(11) & 1.255(19) &
        1.586(36) \\
    $28^3\times 32$ & 118 & 3.5 & 0.01/0.05 & 0.0138 & 0.364(1) & 0.387(7) &
        1.196(25) & 1.561(41) \\ \hline
$20^3\times 64$ &200  & 2.5 &  0.03/0.05  &0.0478 & 0.594(1) &0.585(7) &1.416(20) & 1.683(22)\\
$20^3\times 64$ & 198& 2.5 & 0.02/0.05 &0.0313 & 0.498(3) &0.525(8) &1.261(17) & 1.589(35)\\
$20^3\times 64$ &100 & 2.5 & 0.01/0.05 &0.0138 & 0.362(5) & 0.401(13)&1.139(25) & 1.488(71)\\
$28^3\times 64$ &300 & 3.5 & 0.01/0.05 &0.0138 & 0.353(2) & 0.368(8)&1.191(19) & 1.533(27)\\
\hline \hline 
\end{tabular}
  \caption{Parameters and number of gauge field configurations used
    for the hybrid action.}
  \label{Table:Hybrid params}
\end{table}

In the hybrid action approach  we have applied 
hypercubic (HYP)-smearing~\cite{HYP:2001} to the gauge fields.
 We have 
performed the computation using two different
boundary conditions (b.c.) in the temporal direction.
In the first case we impose Dirichlet b.c. on time slices 0 and 31 
using only  the first half of the lattice and in the second we 
  use the full lattice with
antiperiodic b.c. consistent with the b.c. used in  the
production of the gauge fields.
 The lattice spacing  $a=0.1241$ fm has been determined from
heavy-quark spectroscopy~\cite{Aubin:2004wf} with a statistical
uncertainty of $2\%$.  
For Wilson fermions we use antiperiodic b.c.

To improve convergence to the nucleon and $\Delta$ ground states we
use Gaussian or Wuppertal smearing  to create smeared quark fields as
described in Refs.~\cite{Alexandrou:axial2007prd,Alexandrou:EM_N2006prd} with smearing parameters
$\alpha= 4$ and $n=50$.

The parameters and number of gauge field
configurations are summarized in 
Table~\ref{Table:Wilson params} for Wilson fermions and in 
Table~\ref{Table:Hybrid params} for the hybrid action.

Domain wall
fermions (DWF)~\cite{Kaplan:1992bt,Shamir:1993zy,Narayanan:1992wx} introduce
an additional fifth dimension  of length  $L_5$. 
They preserve the Ward-Takahashi
identity~\cite{Furman:1994ky} even at finite lattice spacing in the
limit $L_5\to\infty$. At sufficiently small values of the lattice
spacing $a$, the effect of a finite value of $L_5$ can be parameterized
by an additional   residual mass term in the Ward-Takahashi
identity~\cite{Blum:2000kn,Blum:1998ud}. This behavior describes a
residual explicit breaking of chiral symmetry which can be minimized
by choosing a sufficiently large extra dimension, $L_5$. We have found
that a value of $L_5=16$ in lattice units 
is sufficient to keep the residual mass,
$(am)_{\mbox{\tiny res}}$, at most one order of magnitude smaller than
the domain wall quark mass, $(am)_q^{\mbox{\tiny DWF}}$. 
The height parameter of
the domain wall action has been chosen to be $am_0=1.7$.

The DWF quark masses displayed in Tables~\ref{Table:Hybrid params} have been
tuned by adjusting the lightest pseudoscalar meson in the Asqtad
calculation~\cite{Bernard:2001MILC} to have the same mass as the
pseudoscalar meson using domain-wall fermions. For technical details
of this tuning procedure, see~\cite{Renner:2004,Hagler:2007}.

%
%

\section{\label{sec:plateaus} Extraction of form factors
from lattice results}
In this section we discuss several technical issues,
beyond the general methodology described in the previous section,
that must be addressed before  the N to $\Delta$ form
factors can be extracted reliably from lattice measurements.

\begin{figure}[h]
\begin{minipage}{8cm}
\epsfxsize=8truecm
\epsfysize=10truecm
 \mbox{\epsfbox{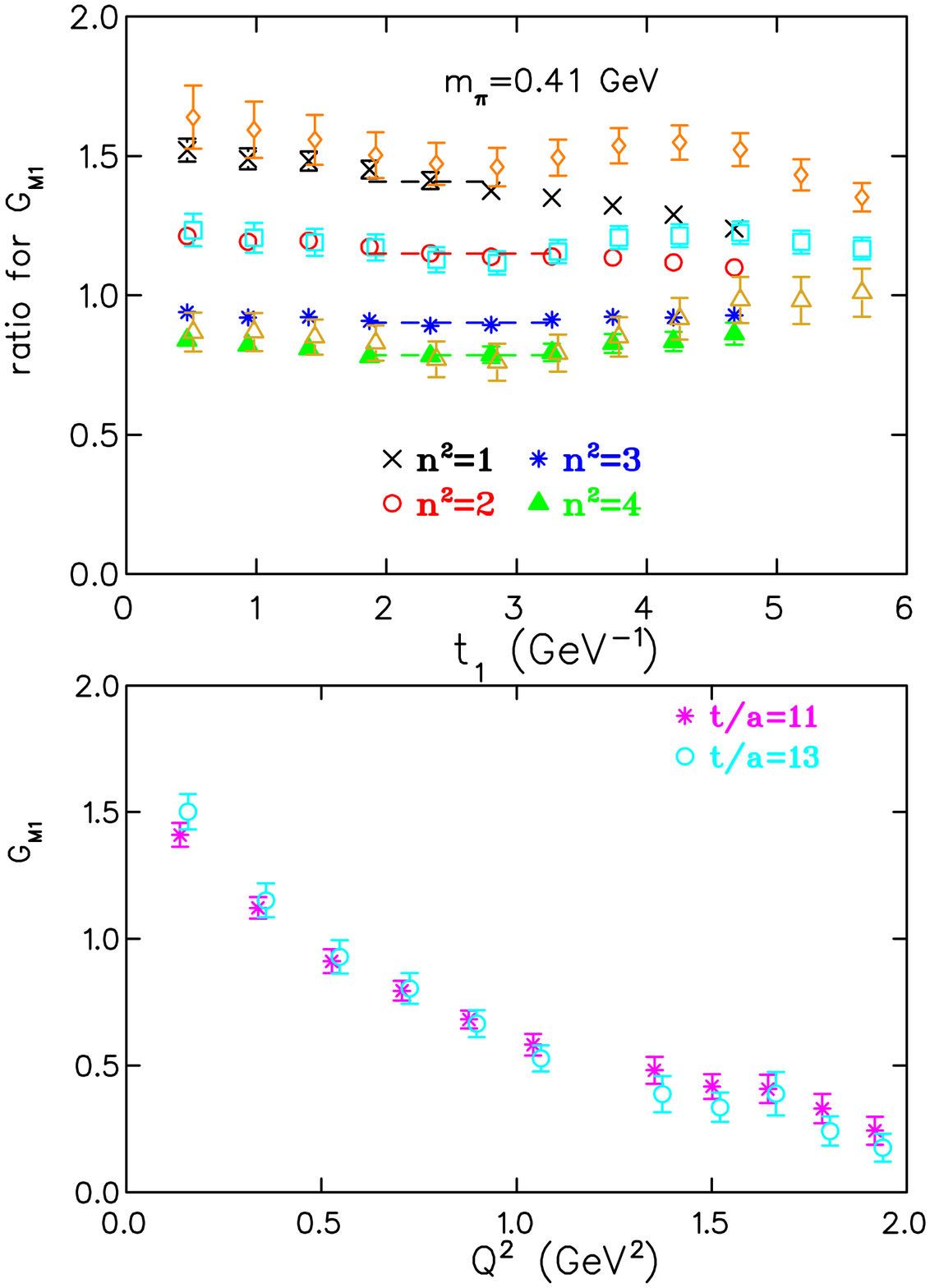}}
\caption{Upper graph: The ratio from which $\GmO$ is
extracted as a function of time separation for the four lowest
$Q^2$-values taking the sink-source separation $t_2=11a$ 
in the quenched theory at pion mass $m_\pi=411$~MeV. 
 We also show results obtained
with sink-source separation $t_2=13a$,
at the two lowest values of $Q^2$ (rhombus and open squares respectively)
 and  at the fourth lowest (open triangles),
  displaced to the left
for clarity.
The  dashed  lines are fits
to the plateaus obtained with  $t_2=11a$ 
and span the range of fitted points.
The lower graph shows $\GmO$ at $m_\pi=411$~MeV,
for sink-source separations $t_2=11a$ 
(asterisks) and $t_2=13a$ (open circles),  displaced to the right,
for clarity.}
\label{fig:plateaus GM1 quenched}
\end{minipage}
\hfill
\begin{minipage}{8cm}
\vspace*{-0.5cm}
\epsfxsize=8truecm
\epsfysize=10truecm
\mbox{\epsfbox{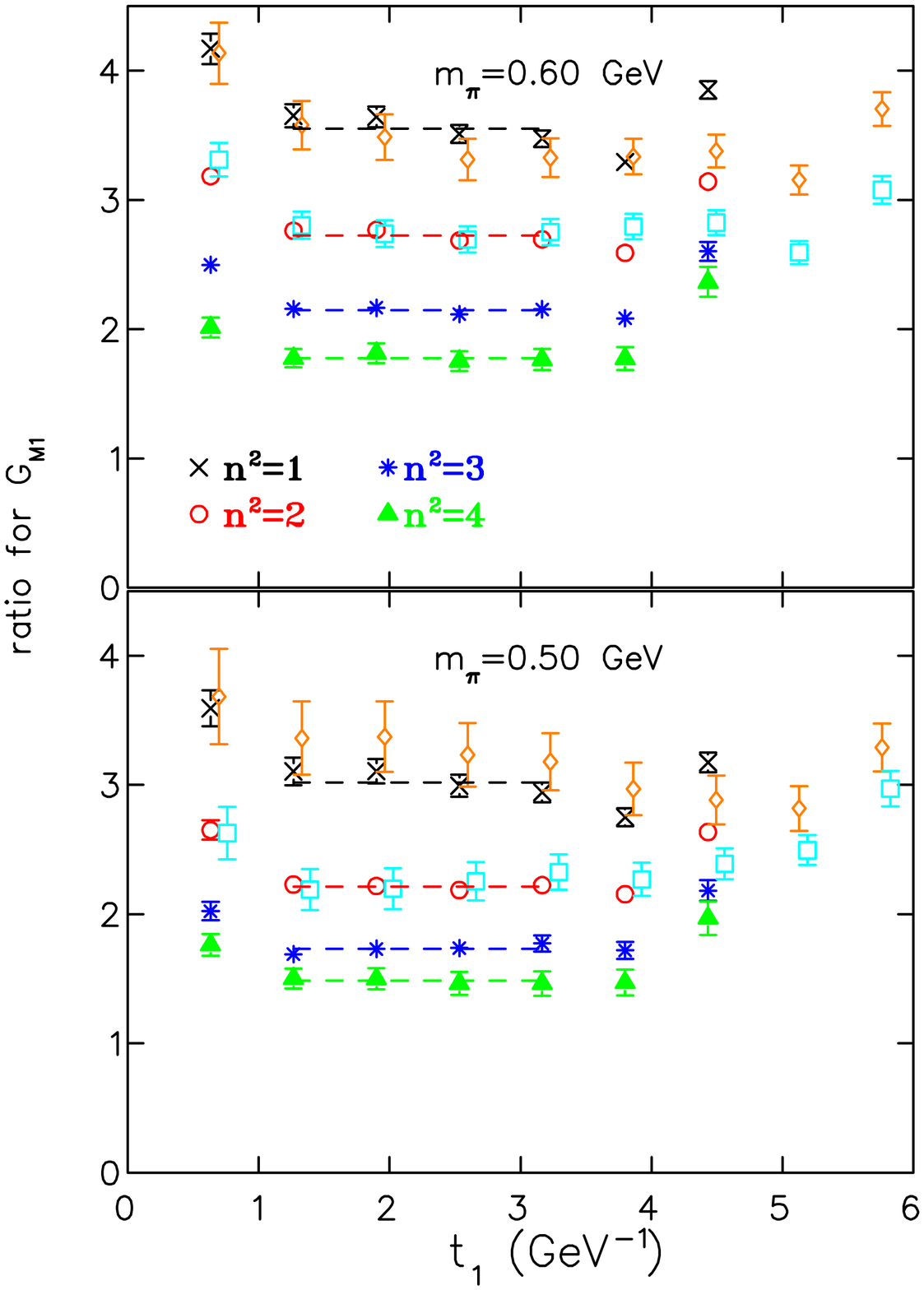}}
\caption{ The ratio from which $\GmO$ is
extracted   for he hybrid  action
using a spatial volume $20^3$ as function of time separation.
We show results for the four lowest
$Q^2$-values taking the sink-source separation $t_2=8a$. 
For the two lowest values of $Q^2$, we show results obtained
with sink-source separation $t_2=10a$(rhombus for
lowest and open squares for second lowest momentum),   displaced to the right,
for clarity.
The upper graph is for $m_\pi=0.594(1)$~GeV and 
the lower for $m_\pi=0.498(2)$~GeV. 
The dashed lines are fits
to the plateaus obtained with  $t_2=8a$ 
and span the range of fitted points.
}
\label{fig:plateaus GM1}
\end{minipage}
\end{figure}

As mentioned already, 
given that sequential inversions are the most time consuming part
of the calculation, the method of choice 
to calculate form factors that are functions of
the momentum transfer squared  is to
perform sequential inversions through the sink. 
However, 
this approach requires that we fix the initial and final hadron states
as well as the sink-source time separation, $t_2$. Changing $t_2$
requires a new sequential inversion.

\begin{figure}[h]
\begin{minipage}{8.cm}
\epsfxsize=8truecm
\epsfysize=5truecm
\mbox{\epsfbox{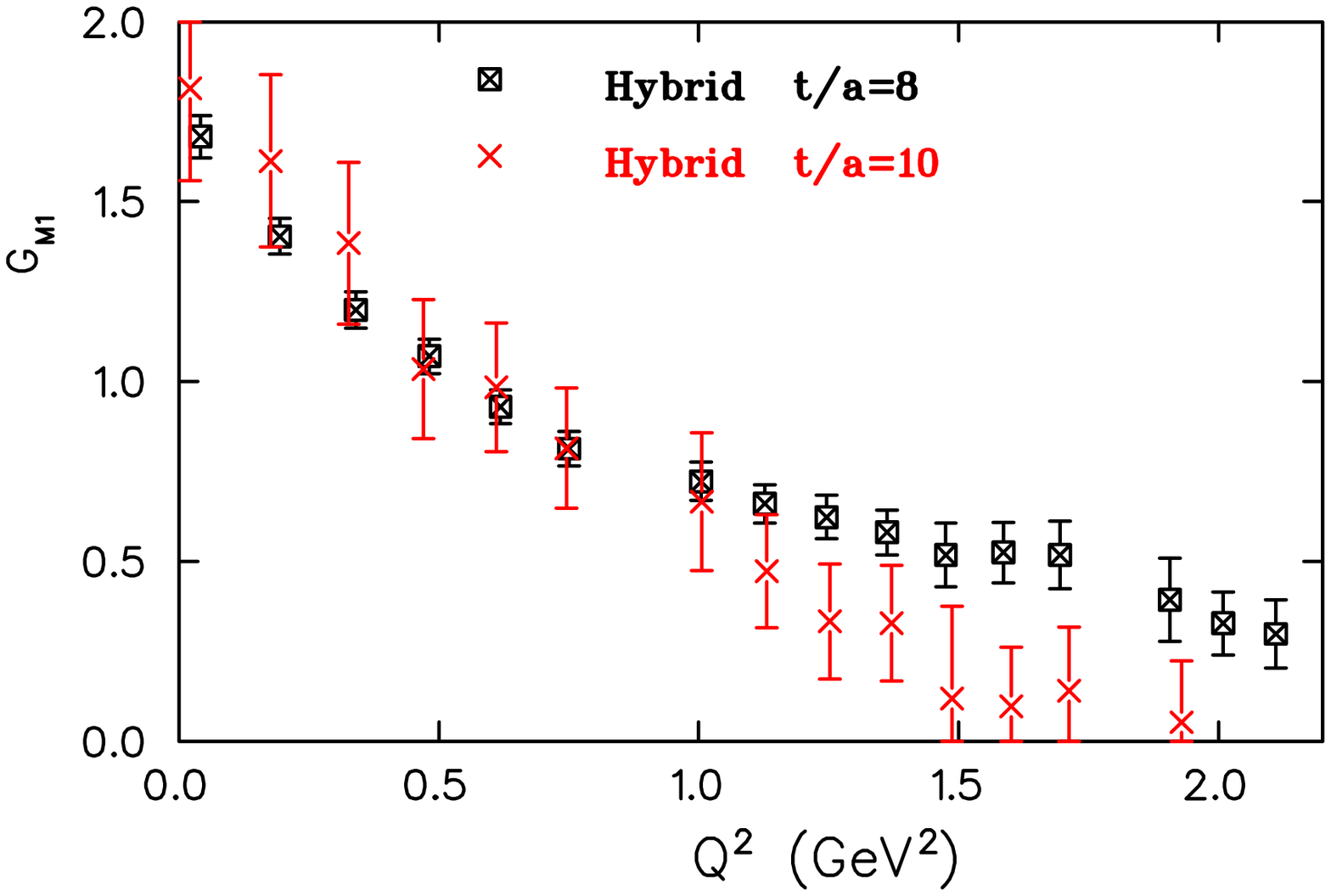}}
\caption{$\GmO$  as function of $Q^2$ at the lowest 
quark mass  for the hybrid   action
for sink-source separations  $t_2/a=8$ and $t_2/a=10$. 
}
\label{fig:GM1 t2}
\end{minipage}
\hfill
\begin{minipage}{8cm}
\epsfxsize=8truecm
\epsfysize=5truecm
 \mbox{\epsfbox{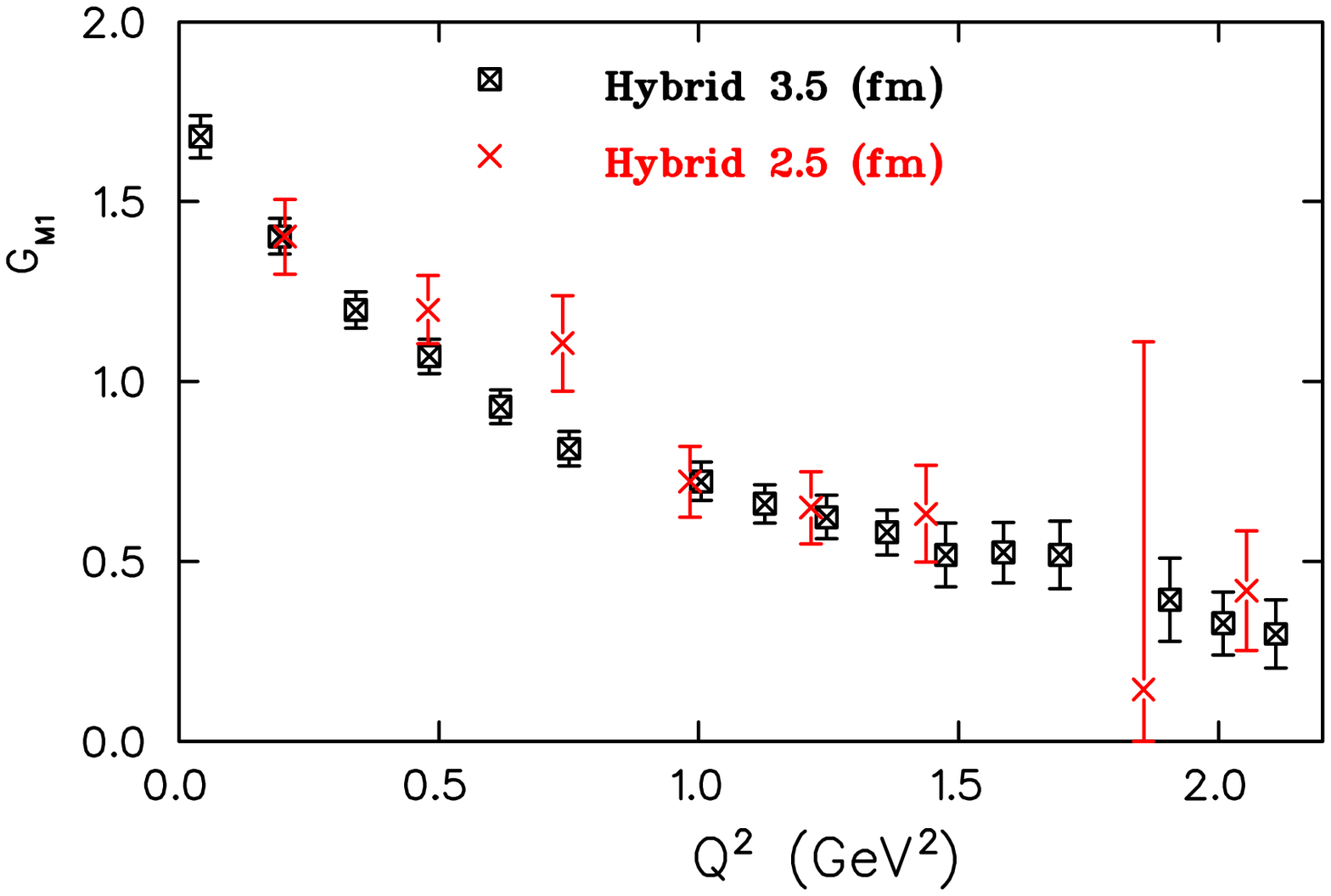}}
\caption{ $\GmO$ as function of $Q^2$ at the lowest 
quark mass  for  the hybrid  action 
for spatial volume $20^3$ and $28^3$. 
}
\label{fig:GM1 vol}
\end{minipage}
\end{figure}
Therefore
we must first determine the optimal value of $t_2$. The criterion is
to choose $t_2$ as small as possible so that statistical errors 
due to the exponential decrease of the signal are minimized but large
enough so that we ensure that excited states
with the nucleon and $\Delta$ quantum numbers are suppressed. 
In order to decide on the value of $t_2$
we compare results at two different sink-source separations
in the quenched theory  and
in the hybrid approach.
 The quenched theory is used to provide a guideline because 
 the statistical errors
are the smallest and therefore deviations due to contamination from excited
states can be seen more easily. We use
two time separations, namely  $t_2/a=11$ and $13$ and
perform the calculation at the smallest pion mass. 
In Fig.~\ref{fig:plateaus GM1 quenched} we show results 
for the ratio $R_{S_1}$ defined
as in Eq.~(\ref{R-ratio}) but using the optimal source of Eq.~(\ref{S1})
for the three-point function and normalized with the appropriate
kimematical factors such that the plateau value yields $\GmO$.
Note that this is not what is actually done in the overconstrained analysis
where the plateau value for each value of the momentum vector is
extracted. Nevertheless $R_{S_1}$ shown in 
 Fig.~\ref{fig:plateaus GM1 quenched}
gives an idea of the quality of the plateaus that are obtained.
We show results for the four lowest $Q^2$-values 
when the sink-source separation  is taken $t_2/a=11$. 
We compare the results obtained  for sink-source separation $t_2/a=13$
at the two lowest 
$Q^2$ values and at the largest value shown in the figure.
The third lowest $Q^2$ value corresponding to $n^2=3$  
for $t_2/a=13$ is
not included since it will make the figure difficult to read. 
As can be seen
the two time separations yield consistent results over a time range
$t_1$. Although in this comparison
we use the same number of configurations
 the statistical errors are much smaller for the shorter
time separation. Fitting to a constant over the plateau range but now
for individual momentum vectors within the overconstrained analysis
we obtained $\GmO$. The results are 
shown in Fig.~\ref{fig:plateaus GM1 quenched}, for 
the two different time separations using 100 configurations in each case.
As can be seen there is very good agreement showing that a time separation of
about 5~GeV$^{-1}$ is sufficient.

 It is important to ensure
that what we find in the quenched case carries over to the hybrid action.
We expect pion cloud effects to be important for dynamical
quarks and therefore we must check that the time evolution is large enough
to allow the pion cloud to fully develop. Guided by our findings in the 
quenched theory we choose
$t_2/a=8$ or about 5~GeV$^{-1}$ and $t_2/a=10$ or about 
6.3~GeV$^{-1}$. In Fig.~\ref{fig:plateaus GM1} we show the ratio $R_{S_1}$
for the two heaviest quark masses whereas in Fig.~\ref{fig:GM1 t2} we 
show the results for $\GmO$ at the smallest quark mass for the
two different time separations. As can be seen,
 $R_{S_1}$  at the
two lowest $Q^2$ values yields the same plateau value
for both time separations. For  $t_2/a=10$ the
statistical errors  are larger
associated with the larger time separation.  The values extracted for
$\GmO$ at the smallest pion mass, where pion cloud effects are
expected to be the largest, are also consistent for the
two time separations, as can be seen in Fig.~\ref{fig:GM1 t2}. The
deviations observed at $Q^2>1.5$ are due to the large statistical
noise associated with the larger time evolution. We therefore conclude
from this analysis that a time separation of about  5~GeV$^{-1}$ is 
also sufficient for our unquenched study.
As a result we fix $t_2\sim 5 $~GeV$^{-1}$ or in lattice units
 to $11$ in the quenched case, to $12$ for
 $N_F=2$ dynamical Wilson fermions and to $8$ in the hybrid approach.

Another potential source of a systematic error
is the spatial size of our lattices.
Given that for the quenched case we use a lattice of spatial size
of about 3~fm
we  expect finite volume effects to be negligible.
Since we do not
have dynamical Wilson configurations on a larger volume we test for
finite size effects in the hybrid scheme for which,  at the smallest quark
mass, there are
 MILC configurations for spatial lattice size $L_s=2.5$~fm and $L_s=3.5$~fm
 giving $L_sm_\pi=4.6$
and $L_s m_\pi=6.4$, respectively.
In Fig.~\ref{fig:GM1 vol}
 we show results for $\GmO$
for these two spatial
sizes.
Results on the smaller lattice are consistent with results on the larger
lattice. This indeed shows that finite volume effects
are small for  $L_sm_\pi\stackrel{\sim}{>}4.5$.  Since for all our
quark masses, except the lightest mass dynamical
Wilson fermions
 $L_sm_\pi>4.6$, we expect finite volume effects to be small.

\section{\label{sec:results-nucleon-tran}Results for the nucleon
  to $\Delta$ transition form factors}

\begin{figure}[h]
\begin{minipage}{8.cm}
\epsfxsize=8truecm
\epsfysize=10truecm
 \mbox{\epsfbox{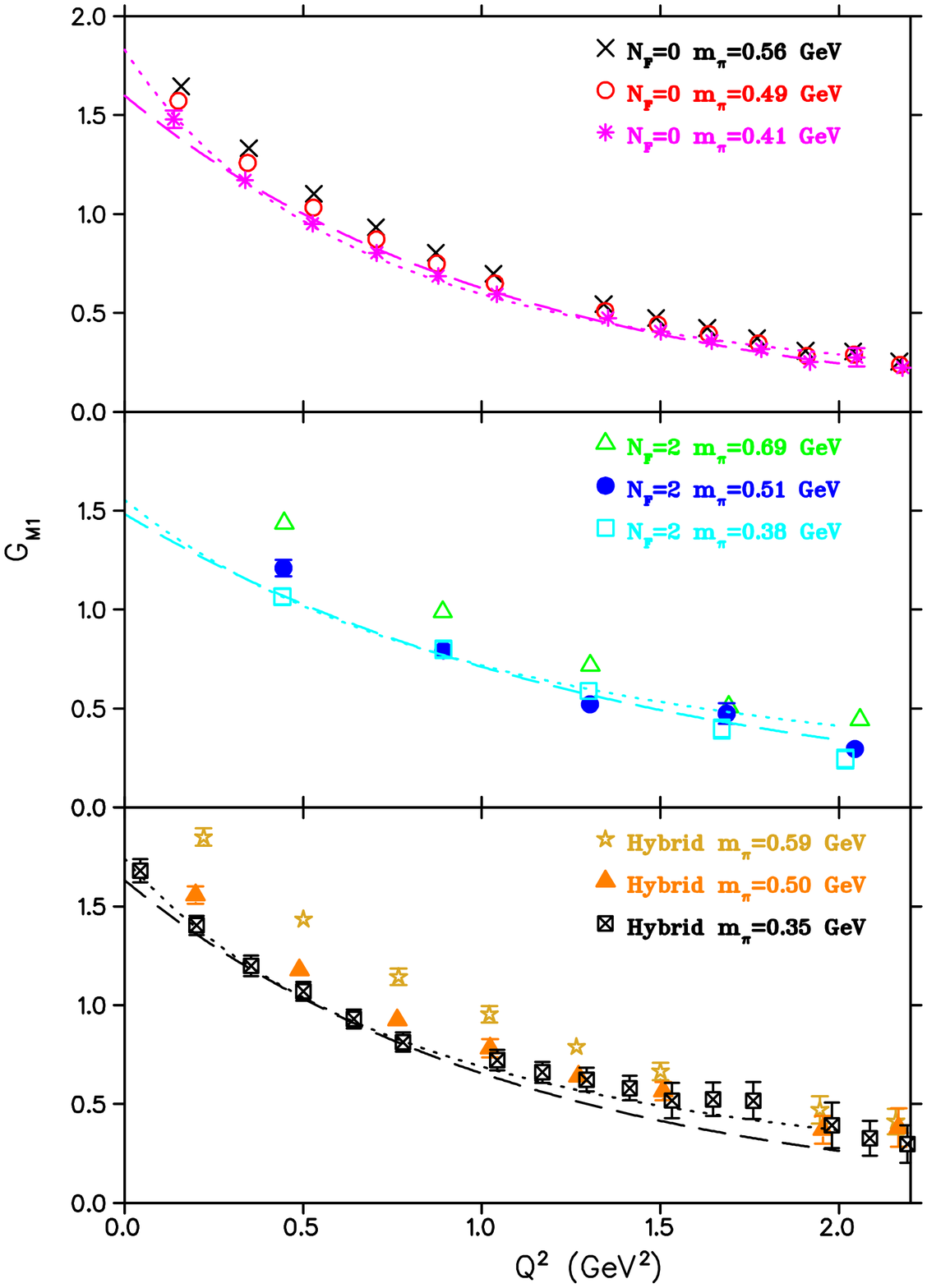}}
\caption{The dipole form factor $\GmO$ as a function of $Q^2$.
The upper graph is for the quenched theory where
results at $m_\pi=563$~MeV are denoted by the
crosses, at $m_\pi=490$~MeV  by the open circles and at $m_\pi=411$~MeV by the
asterisks. 
The middle graph is for dynamical
Wilson fermions 
where results at $m_\pi=691$~MeV are denoted by the
open triangles, at $m_\pi=509$~MeV  by the filled circles and at 
$m_\pi=384$~MeV by the open squares.
The lower is for the hybrid action,
where results at $m_\pi=594$~MeV are denoted by the
stars, at $m_\pi=498$~MeV  by the filled triangles  and at 
$m_\pi=353$~MeV by the inscribed squares. 
The dashed line is a fit to an exponential form $f_0 \exp(-Q^2/m^2)$ and
the dotted line  is a fit to a dipole form $g_0/(1+Q^2/m^2_0)^2$ 
at the lowest pion
mass in each of the three types of simulations.}
\label{fig:GM1_3sim}
\end{minipage}
\hfill
\begin{minipage}{8.cm}
\vspace*{-0.5cm}
\epsfxsize=8truecm
\epsfysize=10truecm
 \mbox{\epsfbox{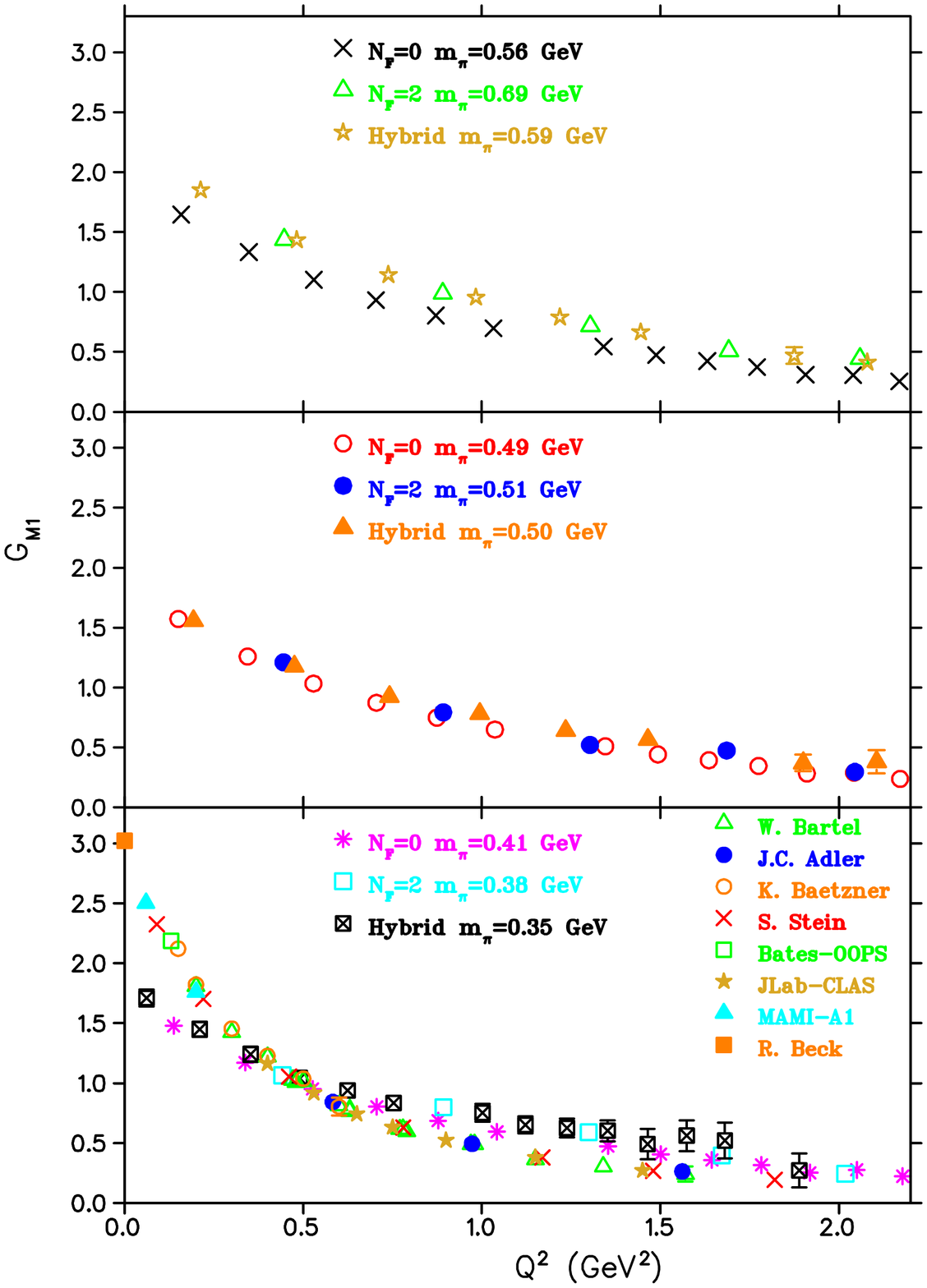}}
\caption{The dipole form factor  $\GmO$ as a function of $Q^2$.
The upper graph is for the heaviest quark mass, the middle graph for
the intermediate quark mass and the lower for the lightest quark mass
for the three type of simulations considered in this work. The notation
for the lattice results is the same as that in Fig.~\ref{fig:GM1_3sim}.
The lattice results at the smallest pion   mass  (lower graph)
 are compared to experiment where  open triangles are results from 
Ref.~\cite{Bartel:1968}, filled circles from Ref.~\cite{Alder:1972}, open
circles from Ref.~\cite{Baetzner:1972}, crosses from Ref.~\cite{Stein:1975},
open squares from Ref.~\cite{Mertz:2001}, stars 
from Refs.~\cite{Joo:2001,Tiator:2001, Tiator:2003}, filled triangles 
from Refs.~\cite{Stave:2006,Sparveris:2007}
and the filled square at $Q^2=0$ from Ref.~\cite{Beck:2000}.
}
\label{fig:GM1_3qmass}
\end{minipage}
\end{figure}

The lattice results  for the dominant dipole form factor are shown in 
Fig.~\ref{fig:GM1_3sim} as we change the quark mass in the three 
types of simulations considered in this work, namely in the quenched
theory denoted by $N_F=0$, for two degenerate flavors of dynamical 
Wilson fermions, denoted by $N_F=2$ and in the hybrid scheme. 
They are also given 
in Tables~III, IV and V
of the Appendix.
All the results discussed in this section in the hybrid approach 
are obtained on the lattices of temporal extent $64$ and
using antiperiodic boundary conditions in the temporal direction.
  
In all three cases the magnetic dipole decreases 
with the quark mass. The dashed lines are
fits to an exponential Ansatz $f_0\exp(-Q^2/m^2)$ whereas
the dotted lines to a dipole Ansatz $g_0/(1+Q^2/m_0^2)$
at the lightest quark mass.
As can be seen, both provide a good description to
the lattice results. The dipole mass that we find in the hybrid approach
at the smallest pion mass is $m_0=1.30(3)$~GeV. For comparison, 
a dipole fit to the experimental results yields $m_0=0.78$~GeV reflecting
the faster fall off of the experimental results.
In Fig.~\ref{fig:GM1_3qmass} we compare quenched and
unquenched results for $\GmO$ at the three quark masses. For similar
pion mass the results are in agreement even for pion mass as low as 350~MeV.
As can be seen, lattice results fall approximately on  the 
same curve having a weaker $Q^2$ dependence than the experimental 
results. In the momentum range considered here experimental results
 can be well described by a dipole form.
In fact, whereas the experimental results fall off faster than the dipole
form factor of the nucleon $G_D=1/(1+Q^2/0.71)^2$ with $m_0=0.78<\sqrt{0.71}$,
 the lattice results
display a weaker $Q^2$ dependence yielding larger values for $m_0$. 
It remains an open question whether decreasing
the quark mass towards  the physical limit will modify this
$Q^2$ dependence.

\begin{figure}[h]
\begin{minipage}{8cm}
\epsfxsize=8truecm
\epsfysize=5truecm
 \mbox{\epsfbox{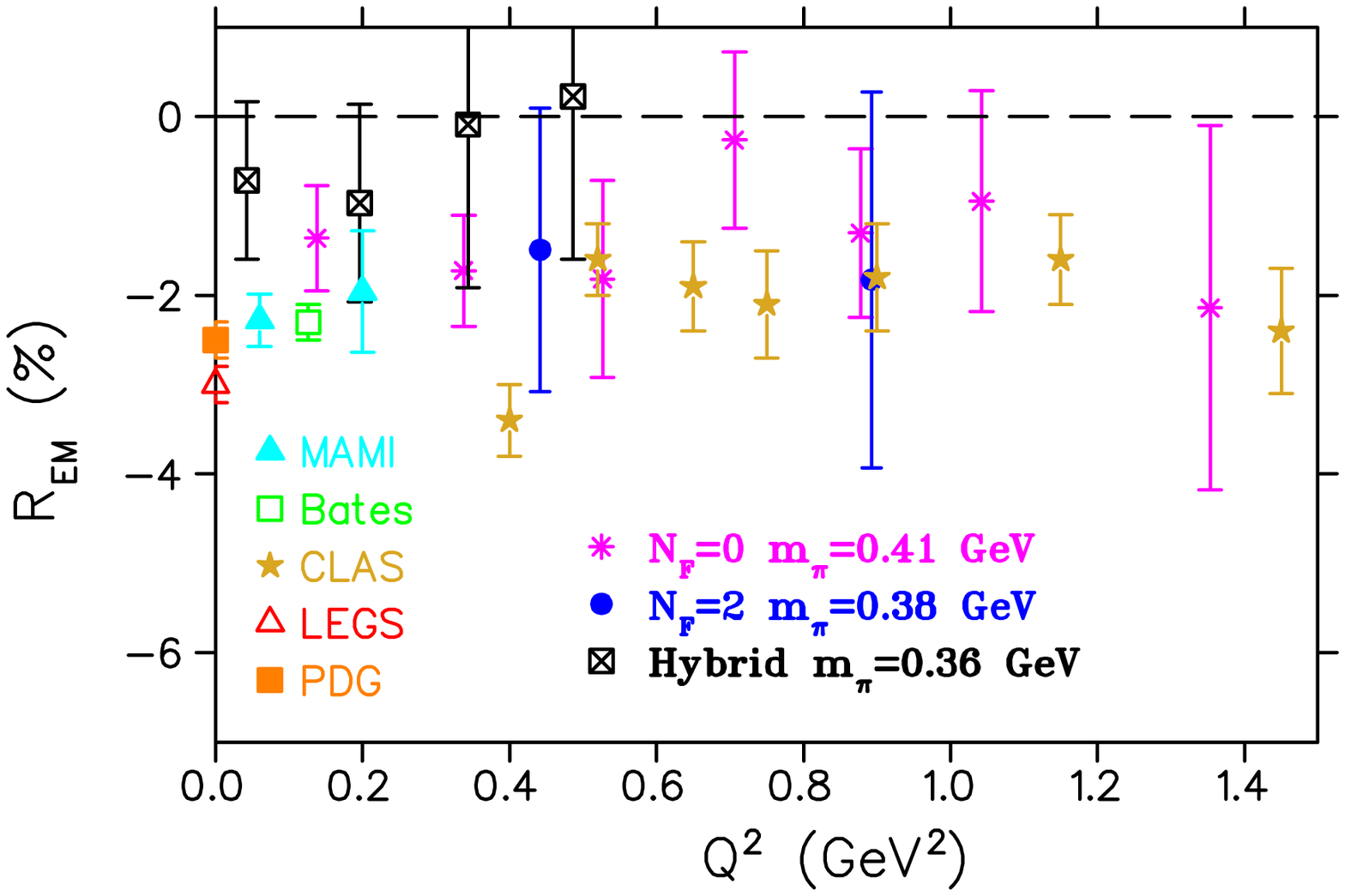}}
\caption{ Lattice calculations of EMR, denoted by $R_{EM}$, for the lightest quark mass  with each action.
 We include experimental
results on EMR from Refs.~\cite{Sparveris:2007, Stave:2006} (filled triangles),
 \cite{Mertz:2001} (open square), \cite{Joo:2001} (stars),
\cite{Blanpied:1996}(open triangle) and \cite{PDG:2002} (filled square).
}
\label{fig:EMR light}
\end{minipage}
\hfill
\begin{minipage}{8cm}
\epsfxsize=8truecm
\epsfysize=5truecm
 \mbox{\epsfbox{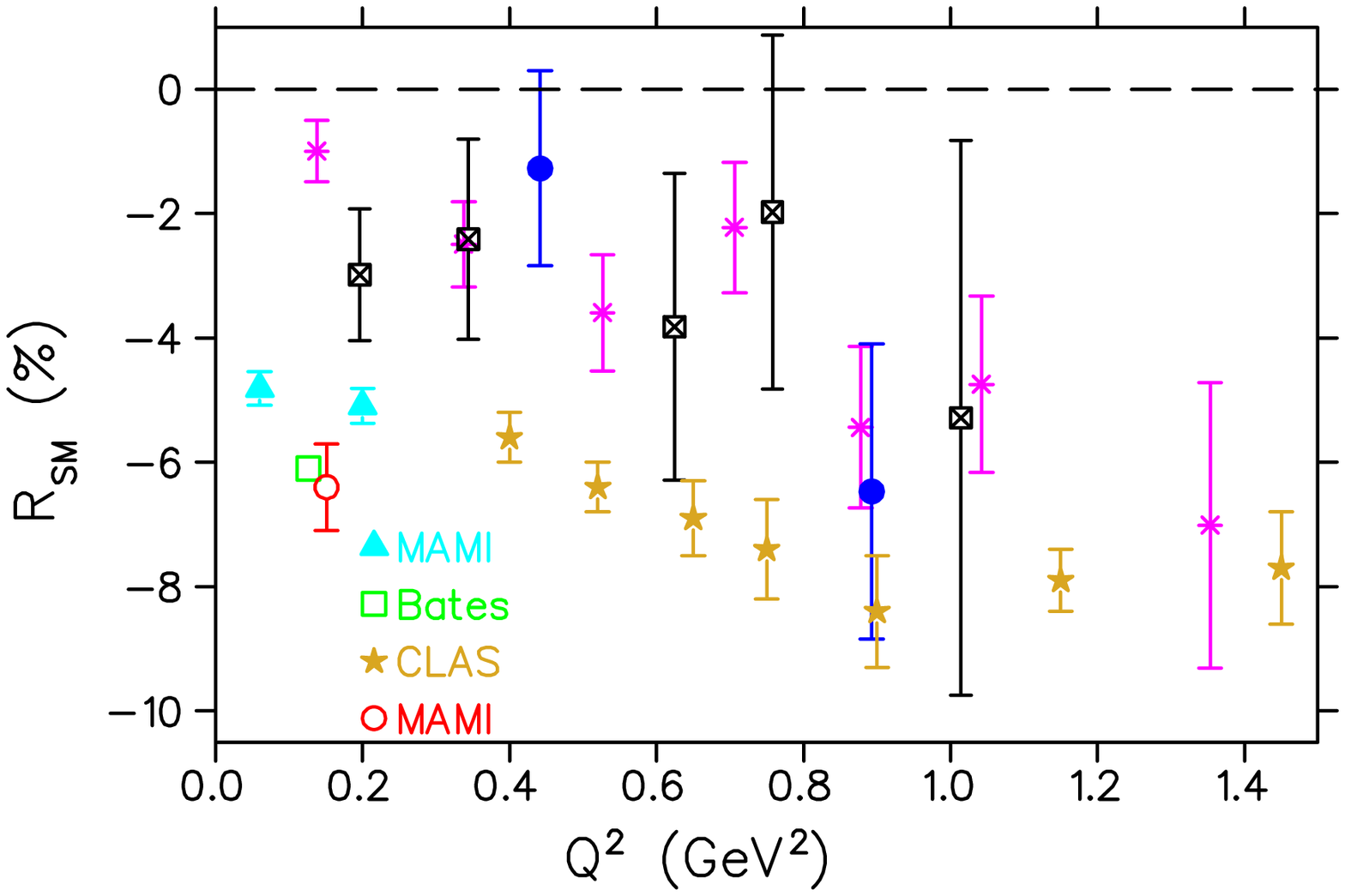}}
\caption{  Lattice calculations of CMR for the lightest quark mass 
 with each action. We include experimental
results on CMR. The notation is the same as that of Fig.~\ref{fig:EMR light}.
The open circle shows the result from Ref.~\cite{MAMI2:2001}.
}
\label{fig:CMR light}
\end{minipage}
\end{figure}

In Figs.~\ref{fig:EMR light} and \ref{fig:CMR light} we show the
results for the EMR and CMR ratios at the lightest quark mass, whereas
we give the complete set of numbers
in Tables~III, IV and V. 
In the case of Wilson fermions we use both sink types
$S_2$ and $S_3$ to extract 
${\cal G}_{E2}(Q^2)$ and ${\cal G}_{C2}(Q^2)$. For domain wall
fermions, for which inversions are very expensive,
we only use $S_2$, which means that ${\cal G}_{C2}(Q^2)$ is not determined
at the lowest momentum transfer as well as at some intermediate values.
However, we obtain results at the second lowest momentum transfer squared
which for the lowest
pion mass on the $28^3$ lattice 
is about the same as the lowest momentum on our quenched lattice, enabling
a  nice comparison.
The main conclusion of this study 
is that these ratios are non-zero and negative. 
As can be seen  in Fig.~\ref{fig:EMR light}, 
there is agreement between the quenched results for EMR and those
obtained within the hybrid approach at the smallest
quark mass. For the CMR ratio there is a notable difference: At the lowest
value of $Q^2$  the quenched results yield a value that it is 
clearly negative but smaller in magnitude 
than experiment. Results in the hybrid approach at the smallest quark mass
are negative and larger in magnitude decreasing
the gap between the lattice value and  experiment. This confirms
that pion cloud contributions are significant at small $Q^2$ modifying
the quenched results. 
This is consistent with the expectation from chiral perturbation 
theory~\cite{Pascalutsa:2005} that pion cloud 
contributions become significant at low $Q^2$. Large
pion contributions  are also needed 
in  phenomenological
approaches~\cite{Tiator:2001,Tiator:2003,Sato:2001}.

\newpage
%
%

\section{\label{sec:conclusions}Conclusions}
We have presented results for  the N to $\Delta$ electromagnetic 
transition form factors calculated within full QCD, using
 two degenerate flavors of dynamical
Wilson fermions,   and  using a hybrid action with staggered
sea quarks and domain wall valence quarks. Comparison of these results
with previous quenched calculations~\cite{Alexandrou:EM_N2D2004prl}
do not reveal large unquenching effects in the case of
the dominant dipole form factor,  $\GmO$, for pion masses
down to about 350~MeV.
 All lattice results for $\GmO$
at these quark masses
show a weaker $Q^2$ dependence than  experiment.
Comparison of results  with the hybrid action
 at two different volumes  indicates that volume effects are not larger
than our statistical errors. 
 Therefore we can not attribute this weaker $Q^2$ dependence
to finite volume effects. Agreement between results using
dynamical Wilson fermions and Domain Wall fermions that have  different finite
lattice spacing systematics suggests that it is unlikely to be
due to discretization errors. However,  we cannot presently 
 extrapolate the lattice results to the continuum limit.
Such an extrapolation would require three different lattice
spacings at similar volumes and quark  masses, which are
currently not available.
  For the EMR and CMR ratios, on
the other hand, unquenched lattice
results begin to show agreement with experiment.
One of the main results  of this
study is the quark mass dependence of the CMR ratio at  low $Q^2$.
 Whereas quenched lattice results  underestimate the magnitude of CMR
at low $Q^2$, results in full QCD
become more negative,
bringing  
 lattice results closer to experiment. 
 This demonstrates the importance of full QCD dynamics in the quadrupole 
form factors as one enters the chiral regime.

 The largest conceptual question as we enter the chiral regime in full QCD, 
is how to fully incorporate the physical effect of the decay of the $\Delta $ 
into a pion and nucleon on the transition form factors.  Even when the pion 
is sufficiently light that the $ \Delta $ could decay in an infinite box, 
the $p$-wave pion in a small box can still be above the decay threshold.
 Naive extrapolation to the chiral limit without including the decay threshold 
could produce qualitative errors in form factors, such as the discrepancy 
presently observed between the lattice and experimental magnetic form factors. 
In the event that chiral perturbation theory is sufficiently accurate 
at and above the $\Delta$ decay threshold, finite volume chiral perturbation 
theory would be an attractive framework for addressing decay channel physics. 
Otherwise, extension of finite volume techniques presently used for
 phase shifts will be required to address form factors. 
This presents an exciting and important research challenge.
%

%
%

\begin{acknowledgments}
We would like to thank B. Orth,  Th. Lippert and K. Schilling~\cite{TchiL}
as well as  C. Urbach, K. Jansen, A. Shindler
and U. Wenger~\cite{Urbach} and the MILC collaboration
 for providing the unquenched
configurations used in this work, as well as the LHP collaboration for
providing forward propagators~\cite{Renner:2004}.
A.T. would like to acknowledge support by the University of Cyprus 
and the program
``Pythagoras'' of the Greek Ministry of Education and
G. K. by the Cyprus Research Promotion Foundation.
 This work is
supported in part by the  EU Integrated Infrastructure Initiative
Hadron Physics (I3HP) under contract RII3-CT-2004-506078, 
by the DFG (Forschergruppe Gitter-Hadronen-Ph{\"a}nomenologie), by the
U.S. Department of Energy (D.O.E.) Office of Nuclear Physics under contract
DE-FG02-94ER40818 and by
the National Science Council of Taiwan under the grant 
numbers NSC96-2112-M002-020-MY3 and NSC96-2811-M002-026.

This research used computational resources provided by  the IBM machine at NIC, J\"ulich,
Germany, by  the National Energy Research Scientific
Computing Center, which is supported by the Office of Science of the U.S.
Department of Energy under Contract No. DE-AC03-76SF00098, and by
the MIT Blue Gene computer, supported by the DOE under grant DE-FG02-05ER25681.
 \end{acknowledgments}

%
%

\bibliography{N2Delta_ref}

%
%

\begin{table}
\section{Appendix}
\begin{tabular}{cccc}
\hline \hline\multicolumn{4}{c}
{Wilson fermions}\\
\hline 
$Q^2$ (GeV$^2$) & $\GmO$ & EMR \% &CMR \%\\ \hline
\multicolumn{4}{c}{Quenched, $\beta=6.0,~~a^{-1}=2.14(6)$~GeV} \\
\hline  
\multicolumn{4}{c}{$m_\pi=563(4)$ MeV}\\
\hline
  0.158 & 1.646(30)& -0.72(26) & -0.82(26)       \\
  0.348 & 1.332(23)& -0.86(29) & -2.09(39)    \\
  0.530 & 1.102(29)& -0.60(57) & -2.62(52)    \\
  0.704 & 0.933(22)& -0.51(47) & -1.80(60)       \\
  0.871 & 0.804(22)& -0.82(45) & -3.11(77)       \\
  1.033 & 0.698(23)& -0.62(63) & -2.92(84)      \\
  1.341 & 0.545(25)& -1.33(93) & -4.31(1.21)    \\
  1.488 & 0.474(24)&           &                \\
  1.631 & 0.424(26)&           &                 \\
  1.770 & 0.373(25)&           &              \\
  1.906 & 0.309(28)&           &              \\
  2.039 & 0.306(34)&           &              \\
  2.169 & 0.254(28)&           &              \\
  2.420 & 0.202(52)&           &              \\
\hline
\multicolumn{4}{c}{$m_\pi=490(4)$ MeV}\\
\hline
  0.151 & 1.572(33) & -0.93(36) &  -0.92(33)   \\ 
  0.344 & 1.259(31) & -1.18(38) &  -2.33(48)  \\
  0.529 & 1.033(30) & -1.03(76) &  -2.99(66)   \\
  0.705 & 0.873(26) & -0.47(64) &  -1.97(74)    \\
  0.874 & 0.749(24) & -1.05(62) &  -3.87(93)    \\
  1.037 & 0.649(25) & -0.79(83) &  -3.46(1.04)   \\
  1.346 & 0.510(28) & -1.65(1.28)& -5.44(1.54)  \\
  1.493 & 0.441(25) &            &            \\
  1.636 & 0.393(28) &            &            \\
  1.775 & 0.346(28) &            &             \\
  1.910 & 0.283(31) &            &             \\
  2.042 & 0.290(38) &            &            \\
  2.171 & 0.238(30) &            &             \\
  2.420 & 0.185(52) &            &            \\
\hline
\multicolumn{4}{c}{$m_\pi=411(4)$ MeV}\\
\hline
  0.138 &  1.479(44) & -1.36(59)  &    -0.99(50)  \\ 
  0.338 &  1.171(37) & -1.73(62)  &  -2.49(69)     \\
  0.527 &  0.951(35) & -1.82(1.10)&  -3.59(93)      \\
  0.706 &  0.804(32) & -0.26(99)  &  -2.22(1.05)   \\
  0.878 &  0.687(26) & -1.30(94)  &  -5.44(1.30)   \\
  1.042 &  0.595(29) & -0.95(1.24)&  -4.74(1.42)   \\
  1.353 &  0.475(34) & -2.14(2.04)&  -7.01(1.22)    \\
  1.501 &  0.406(28) &            &                \\
  1.644 &  0.359(30) &            &               \\
  1.783 &  0.317(31) &            &               \\
  1.918 &  0.252(35) &            &               \\
  2.050 &  0.276(45) &            &               \\
  2.178 &  0.223(34) &            &               \\
  2.426 &  0.172(57) &            &                \\
\hline
\hline
\end{tabular}
\caption{Quenched results for $\GmO$, EMR and CMR}
\label{Table:quenched}
\end{table}

\begin{table}
\begin{tabular}{cccc}
\hline \hline\multicolumn{4}{c}
{Wilson fermions}\\
\hline 
$Q^2$ (GeV$^2$) & $\GmO$ & EMR \%&CMR \%\\ \hline
\multicolumn{4}{c}{$N_F=2$ Wilson, $\beta=5.6,~~a^{-1}=2.56(10)$~GeV} \\
\hline  
\multicolumn{4}{c}{$m_\pi=691(8)$ }\\
\hline
  0.447 & 1.437(36) &-0.76(43)   & -1.76(69)   \\
  0.891 & 0.989(31) &-0.86(71)   & -3.03(1.04) \\
  1.304 & 0.717(30) &-1.44(1.33) & -0.88(1.34)  \\
  1.691 & 0.509(40) &-2.28(1.42) & -5.75(1.95)  \\
  2.058 & 0.443(38) &-2.08(1.48) & -10.06(2.41)  \\
  2.407 & 0.341(37) &           &              \\
  3.060 & 0.208(55)&           &              \\
\hline
\multicolumn{4}{c}{$m_\pi=509(8)$ }\\
\hline
 0.445 & 1.210(42) & -1.00(1.28) & -0.93(1.42) \\ 
 0.892 & 0.794(32) & -5.26(1.87) & -6.17(2.10)  \\
 1.303 & 0.521(32) & -5.95(4.43) & -3.49(3.44)  \\
 1.685 & 0.474(52) & -5.65(3.47) & -4.02(3.33)  \\
 2.044 & 0.296(32) &             &              \\
 2.384 & 0.211(48) &             &              \\
\hline
\multicolumn{4}{c}{$m_\pi=384(8)$ }\\
\hline
  0.442 & 1.066(43) & -1.49(1.59)& -1.27(1.57) \\ 
  0.893 & 0.798(44) & -1.83(2.10)& -6.47(2.37)  \\
  1.299 & 0.589(37) &            &             \\
  1.671 & 0.396(48) &            &            \\
  2.017 & 0.244(47) &            &            \\
  2.342 & 0.181(41) &            &            \\
\hline
\hline
\end{tabular}
\label{Table:unquenched}
\caption{Unquenched Wilson results for $\GmO$, EMR and CMR.}
\end{table}

\begin{table}
\begin{tabular}{cccc}
\hline \hline\multicolumn{4}{c}
{Hybrid action}\\
\hline 
$Q^2$ (GeV$^2$) & $\GmO$ & EMR \% &CMR \%\\ \hline
\multicolumn{4}{c}{Hybrid action,~~$a^{-1}=1.58$~GeV} \\
\hline  
\multicolumn{4}{c}{$m_\pi=594(1)$ }\\
\hline
  0.213 & 1.850(44)  &  -0.16(52)  &             \\
  0.482 & 1.434(36)  &  -0.45(67)  & -3.52(1.16) \\
  0.738 & 1.143(41)  &  -0.52(1.42)& -3.88(1.73) \\
  0.983 & 0.954(42)  &             &             \\
  1.218 & 0.789(38)  &             & -6.25(2.78) \\
  1.445 & 0.665(45)  &             &             \\
  1.874 & 0.471(68) &             &             \\
  2.079 & 0.413(64) &             &             \\
  2.278 & 0.363(72) &             &             \\
  2.472 & 0.322(89) &             &             \\
  2.660 & 0.262(145) &             &              \\
  2.844 & 0.172(158) &             &              \\
\hline
\multicolumn{4}{c}{$m_\pi=498(3)$ }\\
\hline
  0.191 & 1.557(46) &  -0.243(91)  &  \\
  0.471 & 1.177(38) &  -1.14(1.21) & -1.96(1.45) \\
  0.735 & 0.924(40) &  -0.56(2.53) & -0.20(2.42)\\
  0.985 & 0.783(46) &              &             \\
  1.224 & 0.641(40) &              &          \\
  1.452 & 0.565(47) &              &          \\
  1.883 & 0.371(70)&              &          \\
  2.087 & 0.381(96)&              &          \\
  2.284 & 0.294(172)&              &          \\
  2.476 & 0.260(240)&              &          \\
  2.662 & 0.104(226)&              &          \\
  2.843 & 0.146(120)&              &          \\
\hline
\multicolumn{4}{c}{$m_\pi=353(2)$ } \\
\hline
  0.042(16) & 1.681(59) & -0.72(88)    &            \\
  0.194(14) & 1.404(49)  & -0.97(1.11)  & -2.98(1.06) \\
  0.341(8)  & 1.199(51)  & -0.09(1.824) & -2.41(1.61) \\
  0.482(9)  & 1.070(48)  &              &             \\
  0.619(8)  & 0.930(47)  &              & -3.82(2.47) \\
  0.751(9)  & 0.813(48)  &              &             \\
  1.005(11) & 0.723(53)  &              &             \\
  1.127(16) & 0.660(53)  &              &             \\
  1.246(16) & 0.623(60) &              &            \\
  1.362(23) & 0.581(62) &              &             \\
  1.475(51) & 0.518(89) &              &              \\
  1.586(18) & 0.525(84) &              &              \\
  1.695(35) & 0.518(94) &              &              \\
  1.906(65) & 0.392(116) &              &               \\
  2.009(26) & 0.328(87) &              &               \\
  2.111(13) & 0.298(94) &              &               \\
  2.209(20) & 0.275(128) &              &               \\
  2.306(23) & 0.230(94) &              &               \\
  2.402(34) & 0.222(124) &              &               \\
  2.497(34) & 0.073(108) &              &                \\
  2.682(20) & 0.087(170) &              &               \\
\hline
\hline
\end{tabular}
\label{Table:hybrid}
\caption{Results in the hybrid approach for $\GmO$, EMR and CMR. For the 
smallest quark mass we include the errors in the determination of $Q^2$
since these are substantial for the small values of $Q^2$ allowed on this lattice.}
\end{table}

\end{document}